\documentclass[pra,twocolumn,showpacs,superscriptaddress,
preprintnumbers,amsmath,amssymb,tightenlines,floatfix]{revtex4}
\usepackage{graphicx}
\newcommand{\beq}{\begin{equation}}
\newcommand{\eeq}{\end{equation}}
\newcommand{\bea}{\begin{eqnarray}}
\newcommand{\eea}{\end{eqnarray}}
\newcommand{\ba}{\begin{array}}
\newcommand{\ea}{\end{array}}
\newcommand{\bc}{\begin{center}}
\newcommand{\ec}{\end{center}}

\newcommand{\bml}{\begin{subequations}}
\newcommand{\eml}{\end{subequations}}
\newcommand{\commentout}[1]{{}}
\newcommand{\bk}{{\bf k}}

\newcommand{\adag}{a^\dagger}
\newcommand{\bdag}{b^\dagger}
\newcommand{\gdag}{g^\dagger}
\newcommand{\phidag}{\phi^\dagger}

\newcommand{\half}{\hbox{$\frac{1}{2}$}}

\newcommand{\HC}{{\rm H.c.}}
\newcommand{\eq}[1]{(\ref{#1})}
\newcommand{\etal} {{\it et al.\/}}
\newcommand{\ibid} {{\it ibid. \/}}
\newcommand{\vol}[1]{{\bf #1}}
\newcommand{\comment}[1]{{}}
\begin{document}

\title{Improved efficiency of stimulated Raman adiabatic
passage in photoassociation of a Bose-Einstein condensate}
\author{Matt Mackie}
\altaffiliation{Presently at QUANTOP--Danish National Research Foundation
Center for Quantum Optics, Department of Physics and Astronomy, University of
Aarhus, DK-8000 Aarhus C, Denmark.}
\affiliation{Department of Physics, University of Turku,
FIN-20014 Turun yliopisto, Finland}
\affiliation{Helsinki Institute of Physics, PL 64,
FIN-00014 Helsingin yliopisto, Finland}
\author{Kari H\"ark\"onen}
\affiliation{Department of Physics, University of Turku,
FIN-20014 Turun yliopisto, Finland}
\author{Anssi Collin}
\affiliation{Helsinki Institute of Physics, PL 64, FIN-00014
Helsingin yliopisto, Finland}
\author{Kalle-Antti Suominen}
\affiliation{Department of Physics, University of Turku,
FIN-20014 Turun yliopisto, Finland}
\affiliation{Helsinki Institute of Physics, PL 64, FIN-00014
Helsingin yliopisto, Finland}
\author{Juha Javanainen}
\affiliation{Department of Physics, University of Connecticut,
Storrs, Connecticut, 06269-3046}

\date{\today}

\begin{abstract}
We theoretically examine Raman photoassociation of a Bose-Einstein condensate,
revisiting stimulated Raman adiabatic passage (STIRAP). Due to collisional
mean-field shifts, efficient molecular conversion requires strong coupling and
low density, either of which can bring about rogue photodissociation to
noncondensate modes. We demonstrate explicitly that rogue transitions are
negligible for low excited-state fractions and photodissociation that is slower
than the STIRAP timescale. Moreover, we derive a reduced-parameter model of
collisions, and thereby find that a gain in the molecular conversion efficiency
can be obtained by adjusting the atom-atom scattering length with off-resonant
magnetoassociation. This gain saturates when the atom-atom scattering length is
tuned to a specific fraction of either the molecule-molecule or atom-molecule
scattering length. We conclude that a fully-optimized STIRAP scheme may offer
the best chance for achieving coherent conversion from an atomic to a
molecular condensate with photoassociation.
\end{abstract}


\maketitle

\section{Introduction}

Photoassociation occurs when a pair of colliding
ultracold atoms absorb a
laser photon~\cite{THO87}, thereby jumping from the
two-atom continuum to
a bound molecular state. If the initial atoms are
Bose-condensed~\cite{AND95,DAV95,BRA97,DAL99}, then the subsequent
molecules
will also form a Bose-Einstein
condensate~\cite{JAV99,HEI00,VAR01}.
On the other hand, the process known as the Feshbach
resonance~\cite{FES92}, which we refer to as magnetoassociation,
occurs when two ultracold atoms collide in the
presence of a magnetic field, whereby the spin of one
of the
colliding atoms flips, and the pair might then jump
from the two-atom
continuum to a bound molecular state~\cite{STW76,TIE92}. As with
collective photoassociation, the
so-formed molecules will comprise a Bose-Einstein condensate (BEC) if
the incident atoms are themselves
Bose-condensed~\cite{TOM98,VAB99,YUR99}.

Early theories of collective association accounted only for the
condensates, neglecting noncondensate
modes~\cite{JAV99,HEI00,VAR01,TOM98,VAB99,YUR99}. Rogue~\cite{KOS00,JAV02}, or
unwanted~\cite{GOR01,HOL01,BOH99}, transitions to noncondensate modes occur
because the dissociation of a zero-momentum BEC molecule need
not take the atoms back to the zero-momentum atomic
condensate,
but may just as well
end up creating two noncondensate atoms with equal-and-opposite momenta. Since
the collective condensate coupling scales like the square root of the laser
intensity (Feshbach-resonance width) and the dissociation rate scales
like the intensity (width), rogue
dissociation is
expected to play a dominant role in strong
photoassociation (magnetoassociation)~\cite{KOS00,JAV02,GOR01,HOL01,BOH99}.

Pioneering experiments~\cite{WYN00} with
photoassociation of $^{87}$Rb condensate were found to be
just on the verge~\cite{KOS00} of coherent atom-molecule
conversion. Next-generation Na~\cite{MCK02} and $^7$Li~\cite{PRO03}
experiments were aimed at the strongly interacting regime, and probed the
predicted~\cite{BOH99,KOS00,JAV02} photoassociation rate limit. Meanwhile,
groundbreaking experiments in magnetoassociation of a Na condensate
demonstrated a tunable scattering length~\cite{INO98}, as well strong
condensate losses for sweeps of the magnetic field across
resonance~\cite{STE99}. Subsequent experiments in Feshbach-tuning the scattering
length led to the formation of stable $^{85}$Rb~\cite{COR00} and
$^{137}$Cs~\cite{WEB02} condensates--which are otherwise incondensible in
significant fractions due to uncooperative (zero-field) scattering
lengths~\cite{RUP95}, as well as a controlled collapse of a condensate with
bursts of atoms emanating from a remnant condensate~\cite{DON01}, a
counterintuitive decrease in condensate losses for an increasing interaction
time~\cite{CLA02}, and collective burst-remnant oscillations~\cite{DON02}. It
turns out that rogue dissociation is at the heart of these magnetoassociation
experiments~\cite{KOK02,MAC02b,KOE03,DUI03}. Most recently, short-lived quantum
degenerate molecules have been created by sweeping a BEC across a
Feshbach resonance~\cite{KXU03}.

When it comes down to creating a molecular
condensate, the catch to photoassociation is that it generally occurs to an
electronically-excited state, and the subsequent irreversible losses to
spontaneous decay defeat the purpose of molecular coherence. Adding a second
laser to drive molecular population to a stable
electronic state,
stimulated Raman adiabatic passage
(STIRAP)~\cite{STIRAP} in
photoassociation~\cite{VAR97,JAV98,COM_REP} of a
Bose-Einstein condensate~\cite{MAC00,HOP01,DRU02,MAC03} has been proposed
as a means for avoiding radiative decay. Likewise, combining
STIRAP with {\em near-resonant} magnetoassociation is also a viable
means to avoid spontaneous decay when creating a stable molecular
BEC~\cite{KOK01,HOR02,MAC02,MAC03b}, although the laser
spectroscopy (bound-bound-bound) is different from the usual photoassociation
(free-bound-bound).

The hallmark of \mbox{STIRAP} is the counterintuitive pulse
sequence, which in photoassociation of a BEC
amounts to adjusting the two lasers so that in the
beginning, when the system is mainly atomic condensate,
the strongest coupling is between the
excited-molecular and stable-molecular condensates, while in the end, when
effectively everything is in the target state, the
strongest coupling is between the atomic and excited-molecular condensate.
As the population is transferred between the atomic and
stable-molecular condensates, the state with the larger
population is always weakly
coupled to the electronically-excited molecular condensate, and
the subsequently
low (ideally zero) population reduces (eliminates)
radiative losses. In principle, STIRAP allows for complete
conversion from an atomic to a molecular condensate.

In practice, however, free-bound-bound STIRAP relies
on a superposition that includes the atoms and stable molecules but excludes
the electronically-excited molecules~\cite{MAC00}, and the conversion efficiency
is reduced from unity when this so-called dark state~\cite{ARI76} is disrupted
by collisions between particles. Specifically, mean-field shifts due to
collisions between condensate particles (atom-atom, atom-molecule, and
molecule-molecule) make it difficult to achieve STIRAP by moving the
system off the required two-photon resonance~\cite{HOP01,DRU02}. Such two-photon
frequency shifts lead to a larger excited-state fraction, and thus more losses
to radiative decay, which of course reduces the molecular conversion
efficiency, defeating the purpose of collective conversion. Nevertheless, the
collisional interaction strength is directly proportional to density, and the
conversion efficiency can be improved by treating the density as an optimization
parameter~\cite{MAC03}. Another collision-avoidance option is STIRAP in an
optical lattice with two particles per site~\cite{DAM03}, although this scheme
is not many-body coherent.

The purpose of this Article is to re-investigate Raman
photoassociation of a nonideal Bose-Einstein condensate,
focusing on improving the efficiency of STIRAP from an atomic to a molecular
condensate. In Sec.~\ref{MODEL}, we introduce the model and briefly review the
idea of improving the molecular conversion efficiency by reducing the density.
Because this system is in fact strongly coupled, for low density or otherwise,
and because rogue photodissociation to noncondensate modes can be prominent for
strongly coupled systems, Sec.~\ref{ROGIES} explicitly demonstrates that rogue
transitions are negligible when the excited-state fraction is low and
photodissociation is slow compared to the STIRAP timescale. The main
contribution of this work is given in Sec.~\ref{FESHTUNE}, which illustrates
that the conversion efficiency can be further improved by tuning the
atomic scattering length with off-resonant magnetoassociation. Overall
discussion is given in Sec.~\ref{DISC}.

\section{ Density Tuning STIRAP}
\label{MODEL}

\begin{figure}
\centering
\includegraphics[width=8.0cm]{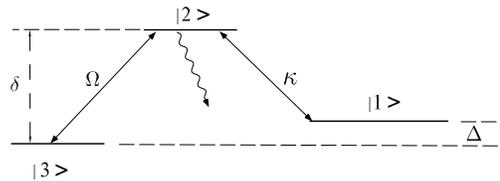}
\vspace{0.5cm}
\caption{
Three-level illustration of coherent free-bound-bound
photoassociation, where $N$ atoms have assumedly
Bose-condensed into state
$|1\rangle$. The free-bound and bound-bound  Rabi
frequencies are $\kappa$ and
$\Omega$, respectively. Similarly, the two-photon and
intermediate detunings
are $\Delta$ and $\delta$. The wavy line denotes the
irreversible losses that STIRAP is intended to manage.}
\label{THREEL}
\end{figure}

Turning to the situation of Fig.~\ref{THREEL}, we
assume that $N$ atoms
have Bose-condensed into the same one-particle state,
e.g., a plane wave
with wave vector $\bk=0$. Photoassociation then removes
two atoms from
this state $|1\rangle$, creating a molecule in the
excited state
$|2\rangle$. Including a second laser, bound-bound
transitions remove
excited molecules from state $|2\rangle$ and create
stable molecules in
state $|3\rangle$. In second quantized notation, boson
annihilation
operators for atoms, primarily photoassociated
molecules, and stable
molecules are denoted, respectively, by $a$, $b$, and
$g$.
The laser-matter interactions that drive the
atom-molecule and
molecule-molecule transitions are characterized by
their respective
Rabi frequencies $\kappa$ and $\Omega$. The two-photon
and intermediate
detunings are denoted by $\Delta$ and $\delta$, and the spontaneous
decay rate of the electronically-excited molecules is $\gamma_s$.

We also introduce atom-atom, atom-molecule, and molecule-molecule contact
interactions with respective coupling strengths:
\bml
\bea
   \lambda_{11} &=& \frac{4\pi\hbar a_{11}}{mV},
\\
   \lambda_{13} &=& \lambda_{31} = \frac{3\pi\hbar a_{13}}{mV},
\\
   \lambda_{33} &=& \frac{2\pi\hbar a_{33}}{mV},
\eea
\eml
where $a_{ij}$ is the particle-particle scattering
length, $V$ is the quantization volume that is expedient in a particular
context (e.g., cubic box or spherical cavity), and
$m$ is the mass of an atom. For the time being, we assume that
the spontaneous decay rate of the
excited-molecular condensate is sufficiently large
to justify neglect of the mean-field
shifts for the electronically excited molecular state
$|2\rangle$. For future reference, we define
$\tilde\lambda_{ij}=V\lambda_{ij}$.

The three-mode Hamiltonian for this freely-interacting system can be written
\bea
\frac{H}{\hbar}&=&\half\Delta \adag a + (\delta-\half i\gamma_s)\bdag b
\nonumber\\
   &&-\half\left[\kappa(\bdag aa + \adag\adag b)
     +\Omega(\gdag b + \bdag g)\right]
\nonumber\\
   &&+\half(\lambda_{11} \adag\adag aa + \lambda_{33} \gdag\gdag gg)
     +\lambda_{13}\adag\gdag ag.
\label{H_0}
\eea
The mean-field equations for collective two-color
photoassociation of a freely-interacting gas are obtained from the
Heisenberg equations of motion:
\bml
\bea
i\dot{a} &=&
  \left(\half\Delta+\Lambda_{11}|a|^2+\Lambda_{13}|g|^2\right) a
    - \chi a^* b, \\
i\dot{b} &=& \left(\delta-\half i\gamma_s \right) b
  -\half\left(\chi aa + \Omega g\right), \\
i\dot{g} &=& \left(\Lambda_{13} |a|^2 + \Lambda_{33} |g|^2\right) g
  -\half\Omega b.
\eea
\label{MFE_0}
\eml
Here the effects of Bose enhancement~\cite{MAC00} have been included by
scaling the mean-field amplitudes according to $x\rightarrow x/\sqrt{N}$ (where
$x=a,b,g$). The mean-field probability $|a|^2$ ($|b|^2,|g|^2$) is then of the
order of unity (half), and the coupling strengths have been redefined as
$\chi=\sqrt{N}\kappa$ and $\Lambda_{ij}=N\lambda_{ij}=\rho\tilde\lambda_{ij}$.
The transient STIRAP pulses are taken as
Gaussian, i.e., $\chi(t)=\chi_0\exp[-\left(t-D_1\right)^2/T^2]$ and
$\Omega(t)=\Omega_0\exp[-\left(t-D_2\right)^2/T^2]$.

Assuming an ideal gas ($\Lambda_{ij}=0$), it is easy to show from the
steady-state limit ($\dot{x}=0$) that Eqs.~\eq{MFE_0} reduce to an algebraic
system that can be plugged trivially into {\it Mathematica}~\cite{WOL96},
yielding exact solutions~\cite{MAC00}. Here we scale the atomic (molecular)
amplitude(s) as $a\rightarrow e^{i\Delta t/2}a$
($b\rightarrow e^{i\Delta t}b$, $g\rightarrow e^{i\Delta t}g$), and keep only
normalized solutions [$|a|^2+2(|b|^2+|g|^2)=1$] that are real
for real values of $\Omega/\chi\rightarrow(0,\infty)$. The dark state is given
as
\bml
\bea
\Delta_0 &=& 0,
\\
a_0 &=& \frac{1}{2}\,\sqrt{\frac{\Omega}{\chi}
  \left[\sqrt{8+\left(\frac{\Omega}{\chi}\right)^2}
    -\frac{\Omega}{\chi}\,\right]},
\\
b_0 &=& 0,
\\
g_0 &=& -\frac{1}{4}
  \left[\sqrt{8+\left(\frac{\Omega}{\chi}\right)^2}
    -\frac{\Omega}{\chi}\right],
\eea
\label{DARKY}
\eml
\commentout{
whereas the ``bright" (nonadiabatic) states, are given as
\bml
\bea
\Delta_\pm &=& \pm\,\frac{\Omega^2+\delta(\delta\mp\sqrt{\delta^2+\Omega^2})}
  {2\sqrt{\delta^2+\Omega^2}}
\\
a_\pm &=& 0,
\\
b_\pm &=& \half\sqrt{1\mp\frac{\delta}{\sqrt{\Omega^2+\delta^2}}},
\\
g_\pm &=& -b_\pm
  \left[\frac{\Omega^2+\delta\left(\delta\pm\sqrt{\Omega^2+\delta^2}\,\right)}
    {\Omega\sqrt{\Omega^2+\delta^2}}\right].
\eea
\label{WHITEY}
\eml
}
It is evident from Eqs.~\eq{DARKY} that, for a two-photon
resonance ($\Delta=0$), the dark state is an eigenstate of the system with all
atoms (stable molecules) for $\Omega/\chi\rightarrow0$
($\Omega/\chi\rightarrow\infty$), and zero excited molecules; hence, for an
initial BEC and adiabatic counterintuitive pulses, the two-photon-resonant system
will follow this eigenstate as it evolves from atoms into stable molecules. The
presence of collisions, which introduce a time-dependent shift of the two-photon
detuning, means that the system cannot start out in, nor subsequently follow,
the dark state, and a nonzero excited-state fraction will ensue.

\begin{figure}[b]
\centering
\includegraphics[width=8.0cm]{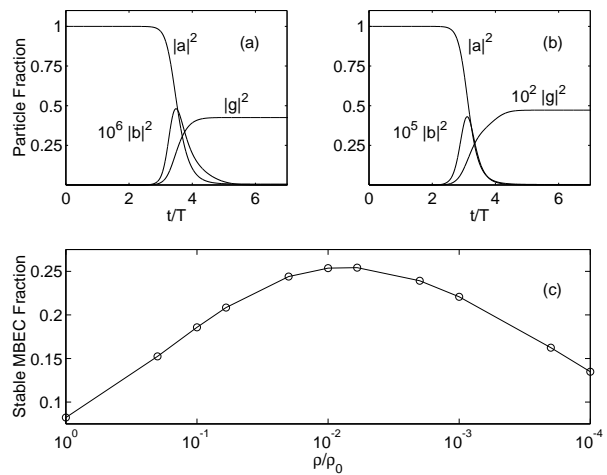}
\caption{Stimulated Raman adiabatic
passage in photoassociation of a nonideal Bose-Einstein
condensate. For a given
$\chi_0(\rho)$, the pulse parameters are $\Omega_0=\chi_0$, $T=5\times
10^3/\chi_0$, $D_1=4.5T$ and
$D_2=2.5T$. The two-photon (intermediate) detuning is
$\Delta=0$ ($\delta=\chi_0$). Note that $|g|^2=1/2$ is actually complete
conversion, since two atoms are destroyed to create a molecule.
(a)~For an ideal gas with $\rho=\rho_0$, STIRAP mostly defeats spontaneous
decay, as expected for a moderately adiabatic pulse sequence. (b)~Adding
collisional interactions shifts the system off resonance, leading to a larger
excited-state fraction and, thus, more losses to spontaneous decay.
(c)~Density-tuning the STIRAP efficiency: Losses to spontaneous decay decrease
as collisional mean-field shifts are marginalized for lower densities, but the
\protect{STIRAP} timescale ($\propto1/\sqrt{\rho}$) increases as well, and
spontaneous decay eventually regains its dominance. Here a larger peak
bound-bound pulse was used,
$\Omega_0=50\chi_0$, as discussed in the text.}
\label{DTUNE}
\end{figure}

To illustrate numerically, we concentrate on explicit parameters for a dilute
quantum-degenerate gas of
$^{87}$Rb atoms~\cite{DRU02}: $\gamma_s=7.4\times 10^7\,\rm{s}^{-1}$,
$\chi_0=2.1\times 10^6\sqrt{\rho/\rho_0}\:\rm{s}^{-1}$,
$\rho_0=4.3\times 10^{14}\,\rm{cm}^{-3}$,
$\tilde\lambda_{11}=4.96\times 10^{-11}\,\rm{cm}^3/s$,
$\tilde\lambda_{13}=-6.44\times 10^{-11}\,\rm{cm}^3/s$; although unknown,
the stable-molecule mean-field shift
$\tilde\lambda_{33}=2.48\times 10^{-11}\,\rm{cm}^3/s$ is estimated by
assuming equal atom-atom and molecule-molecule scattering lengths. The
effect of collisions is demonstrated explicitly in Fig.~\ref{DTUNE}. Setting the
mean-field-shift terms to zero,
$\Lambda_{ij}=0$, we see in Fig.~\ref{DTUNE}~(a) that STIRAP works exactly
as expected for a near-adiabatic pulse sequence, losing only a small fraction of
particles to spontaneous decay. Reinstating collisions, Fig.~\ref{DTUNE}~(b)
shows an order of magnitude increase in the excited-state fraction, and a
corresponding decrease in the conversion efficiency of some two orders of
magnitude, as spontaneous decay gets the upper hand.

Nevertheless, since the collisional mean-field shifts are directly proportional
to the density ($\Lambda_{ij}\propto\rho$), then we expect that reducing the
density of the initial BEC could improve the conversion efficiency of STIRAP
from an atomic to a molecular condensate~\cite{MAC03}. This idea was spawned in
our development of Feshbach-stimulated photoproduction of stable
molecular condensates~\cite{MAC02b} (see also related work~\cite{YUR03}).
Consider for example the density
$\rho=4.3\times 10^{12}\,\rm{cm}^{-3}$, so that
$\chi_0=2.1\times 10^5\,\rm{s}^{-1}$,
$\Lambda_{11}=213\,\rm{s}^{-1}$,
$\Lambda_{33}=107\,\rm{s}^{-1}$,
and $\Lambda_{13}=277\,\rm{s}^{-1}$. The mean-field shifts are
then roughly three orders of magnitude smaller than the peak Bose-enhanced
free-bound coupling
$\chi_0$, and since $\chi_0$ sets the timescale for
atom-molecule STIRAP~\cite{MAC00}, we expect a smaller role for particle
interactions compared to when $\rho=\rho_0$. This intuition is
confirmed in Fig.~\ref{DTUNE}~(c). Note
that pulses with asymmetric heights~\cite{HOP01,DRU02} improve the short-pulse
efficiency of STIRAP by more than an order of magnitude over pulses with
symmetric heights, as can be seen by comparing the final stable fraction in
Fig.~\ref{DTUNE}~(b) with the stable fraction for $\rho/\rho_0=1$ in
Fig.~\ref{DTUNE}~(c). The optimal density for these pulse parameters is
$\tilde\rho\sim 10^{12}\,{\rm cm}^{-3}$, for which conversion to a molecular
BEC occurs on a timescale
$T=5\times 10^3/\chi_0 \sim 50\,\rm{ms}$. Optimizing the detunings and
pulse parameters~\cite{DRU02} should give even further improvement (see also
Ref.~\cite{HOR02}).

\section{Rogue Photodissociation to Noncondensate Modes}
\label{ROGIES}

As mentioned in the introduction, transitions to
noncondensate modes occur because, although formed from a pair of
zero-momentum BEC atoms, a condensate molecule need not photodissociate
back to the atomic condensate, but may just as well form a noncondensate
atom pair with equal and opposite momentum. Although
experiments remain inconclusive on this score~\cite{MCK02,PRO03}, a rate
limit is to be expected when the conversion to rogue-dissociated atom
pairs dominates over the formation of molecular
condensate~\cite{BOH99,KOS00,JAV02}. Similarly, the losses for
across-resonance sweeps~\cite{COR00}, as well as the counterintuitive
losses~\cite{CLA02}, observed in $^{85}$Rb magnetoassociation experiments can be
viewed as collective rapid adiabatic passage from BEC to rogue-dissociated atoms
pairs~\cite{MAC02b}; also,
the remnant-burst oscillations~\cite{DON02} are
identified as Ramsey
fringes in the evolution between an atomic condensate
and a molecular
condensate dressed by noncondensate atom
pairs~\cite{KOK02,MAC02b,KOE03,DUI03}.

The prime indicator of rogue
dominance is the density-dependent frequency
$\omega_\rho=\hbar\rho^{2/3}/m$~\cite{JAV99,KOS00,JAV02}. When the
collective-enhanced atom-molecule coupling satisfies
$\chi_0\agt\omega_\rho$, transitions to noncondensate modes are expected
to dominate atom-molecule conversion. For the
previous example of
$\rho=4.3\times 10^{12}\,\rm{cm}^{-3}$, we find
$\chi_0\sim100\,\omega_\rho$; moreover, even for a typical density
($\rho=\rho_0=4.3\times 10^{14}\,\rm{cm}^{-3}$), we find $\chi_0\sim
50\,\omega_\rho$. It is therefore reasonable to suspect that rogue
photodissociation could play a dominant role in STIRAP from
an atomic to a molecular condensate, for low-density systems or otherwise.

The purpose of this section is to apply our minimal-yet-realistic
model~\cite{JAV02} of rogue dissociation to noncondensate modes in order
to determine its impact on photoassociative STIRAP in a freely-interacting
Bose-Einstein condensate. Accordingly, the Hamiltonian~\eq{H_0} is
adapted to read
\bea
\frac{H}{\hbar}&=&\half\Delta \adag a + \delta\bdag b
   -\half\left[\left(\kappa\bdag aa +\Omega\gdag b\right) + \HC\right]
\nonumber\\
   &&+\half(\lambda_{11} \adag\adag aa + \lambda_{33} \gdag\gdag gg)
     +\lambda_{13}\adag\gdag ag
\nonumber\\&&
  +\half\sum_{\bk\neq0} \left[\left(\Delta+\epsilon_\bk\right)\adag_\bk a_\bk
    -\kappa f_\bk \left(\bdag a_\bk a_{-\bk}
      +\HC\right)\right].
\nonumber\\
\label{H_ROGUE}
\eea
Here $\epsilon_\bk=\hbar k^2/m$ is the energy of a noncondensate pair, and
$f_\bk$ describes the wavenumber (energy) dependence of the noncondensate
coupling. In order to isolate the effects of rogue dissociation, we have
dropped the spontaneous decay term
$i\gamma_s\bdag b$. Also, anticipating low populations, we have
neglected collisions with noncondensate atoms. The corresponding
mean-field theory is derived from the Heisenberg equations of motion:
\bml
\bea
i\dot{a} &=&
  \left[\half\Delta+\Lambda_{11}|a|^2+\Lambda_{13}|g|^2\right]a
    - \chi a^* b, \\
i\dot{b} &=& \delta b-\!\half\!\left[\chi a^2 + \Omega g
  +\xi\!\int\!d\epsilon\sqrt[4]{\epsilon}\,
    f(\epsilon)C(\epsilon)\right]\!\!,\:\:
\\
i\dot{g} &=& \left[\Lambda_{13} |a|^2 + \Lambda_{33} |g|^2\right] g
  -\half\Omega b,
\\
i\dot{C}(\epsilon) &=&\left[\Delta+\epsilon\right]C(\epsilon)
    -\xi\sqrt[4]{\epsilon}\,f(\epsilon)\left[1+2P(\epsilon)\right]b,
\\
i\dot{P}(\epsilon) &=&
  \frac{(2\pi\omega_\rho)^{3/2}}{\sqrt[4]{\epsilon}}
    f(\epsilon)\left[b^*C(\epsilon)-C^*(\epsilon)b\right].
\eea
\label{ROGUE_MFE}
\eml
In Eqs.~\eq{ROGUE_MFE}, we have taken the continuum limit
\beq
\frac{1}{N}\sum_\bk G_\bk \rightarrow
  \frac{1}{4\pi^2\omega_\rho^{3/2}}\int d\epsilon\, G(\epsilon),
\eeq
and we also have introduced the rogue and normal densities,
$C(\epsilon)\equiv\sqrt[4]{\epsilon}\,
  \left\langle a_\bk a_{-\bk}\right\rangle/(2\pi\omega_\rho^{3/4})$
and $P(\epsilon)\equiv\langle \adag_\bk a_\bk\rangle$, as well as
the rogue coupling $\xi(t)=\chi(t)/(2\pi\omega_\rho^{3/4})$.

Before moving on, we discuss basic renormalization. For
$\chi_0=\Omega_0=0$
and ignoring the normal density, simple Fourier analysis of an initial
excited-bound molecular condensate gives the below-threshold binding energy as
the real and negative pole of
\beq
\omega-\delta-\Sigma(\omega)+i\eta=0,
\eeq
where $\eta=0^+$ and the molecular self-energy is defined as
\beq
\Sigma(\omega)=\half\xi^2\int d\epsilon\,f^2(\epsilon)\,
  \frac{\sqrt{\epsilon}}{\omega-\epsilon+i\eta}\,.
\eeq
The simplest energy dependence for the continuum is one that obeys the
Wigner threshold law up to some abrupt cutoff:
$f^2(\epsilon)=\Theta(\epsilon_M-\epsilon)$. The detuning (binding energy) then
picks up a term $\Sigma(0)=\xi^2\sqrt{\epsilon_M}$\,. In principle, the cutoff
is infinite, and therefore so is the continuum shift of the
molecular binding energy. To account for this divergence, one defines the
so-called physical detuning
$\tilde\delta=\delta-\xi^2\sqrt{\epsilon_M}\,$, which is finite by
definition in the limit of an infinite cutoff.

In practice, any numerical
procedure employs a finite cutoff, and the finite shift is accounted for
in exactly the same manner. Hence, the intermediate detunings in
Eqs.~\eq{ROGUE_MFE} are taken as physical (renormalized) detunings. Anticipating
low noncondensate fractions, the normal density--Bose enhancement of the
noncondensate modes--was neglected. Reminiscent of our original nondegenerate
quasicontinuum model of photoassociation~\cite{JAV98}, we introduce
$N_{qc}=2\times10^3$ quantum-degenerate quasicontinuum states and a cutoff
$\epsilon_M=10\,\chi_0$, sufficient to deliver convergence and keep
numerical artifacts to a minimum, where convergence is determined by the
requirement
\beq
|a|^2+2\left[|b|^2+|g|^2\right]+\int d\epsilon\,|C(\epsilon)|^2=1.
\eeq

\begin{figure}[b]
\centering
\includegraphics[width=8.0cm]{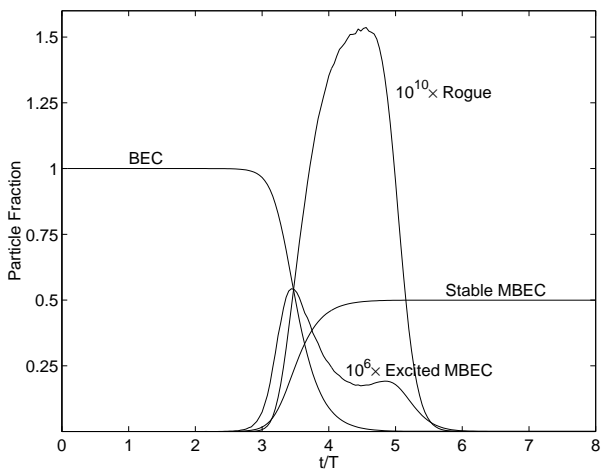}
\caption{Rogue photodissociation to noncondensate modes in stimulated
Raman adiabatic passage in photoassociation of an interacting
Bose-Einstein condensate. The STIRAP parameters are
$\delta=\Omega_0=\chi_0$, $T=5\times10^3/\chi_0$, $D_1=4.5T$, and
$D_2=2.5T$.
The fraction of rogue (noncondensate) particles,
$f_r=\int d\epsilon\,|C(\epsilon)|^2$, is clearly negligible, and STIRAP runs
even better than in the ideal case [see Fig~\protect\ref{DTUNE}~(a)] because
photodissociation is slower than spontaneous decay.}
\label{ROGUE_FIG}
\end{figure}

That said, we report numerical STIRAP solutions of the
mean-field theory~\eq{ROGUE_MFE} for rogue transitions to noncondensate modes
in Raman free-bound-bound photoassociation of a BEC.
The focus is on the lowest density from Fig.~\ref{DTUNE},
$\rho=4.3\times 10^{10}\,\rm{cm}^{-3}$, as an explicit example  in
Fig.~\ref{ROGUE_FIG}, and the key results are summarized in
Table~\ref{ROGUE_TBL}. In this case, not only is the coupling strong to begin
with, but we have an especially low density. Here the
rogue coupling is a bit stronger, due to the low density, leading to a slightly
larger noncondensate fraction. But, due to the fact that
mean-field shifts are effectively eliminated for the lowest
density, the bound state fraction is in this case the lowest. In the end, even
for this low density, we see that rogue photodissociation to noncondensate modes
does not affect the final stable-molecule conversion efficiency. We have
confirmed that, for artificially large rogue couplings $\xi\agt 1$, i.e., faster
photodissociation, rogue noncondensate modes indeed begin to play a more
invigorated role. Also, if the timescale for STIRAP is
lengthened too much, we expect that rogies should likewise begin to dominate.

\begin{table}[t]
\caption{Overall role of noncondensate modes for stimulated Raman adiabatic
passage in photoassociation of an interacting Bose-Einstein
condensate. The STIRAP parameters are as in Figure~\protect\ref{ROGUE_FIG}.
Recall that the density scale is $\rho_0=4.3\times10^{14}\,\rm{cm}^{-3}$, the
rogue indicator is
$\omega_\rho=\hbar\rho^{2/3}/m$, and the fraction of noncondensate particles is
$f_r=\int d\epsilon\,|C(\epsilon)|^2$. Whereas the system is strongly coupled
even for large densities, i.e., $\chi_0\agt\omega_\rho$, the excited-state
fractions are low enough, and rogue dissociation slow enough, to guarantee that
a negligible fraction of the population is transferred to the noncondensate
modes.}
\begin{ruledtabular}
\begin{tabular}{cccccc}
$\rho/\rho_0$ & $\chi_0/\omega_\rho$
  & $f_b)_{\rm max}$ & $f_r)_{\rm max}$ & $f_g$
\\
\hline
1 & $5.05\times10^1$ & $3.45\times10^{-4}$ & $2.41\times10^{-11}$
  & 0.49986
\\
$10^{-2}$ & $1.08\times10^2$ & $3.69\times10^{-6}$ & $1.74\times10^{-11}$
  & 0.49983
\\
$10^{-4}$ & $2.34\times10^2$ & $5.43\times10^{-7}$ & $1.54\times10^{-10}$
  & 0.49952
\end{tabular}
\end{ruledtabular}
\label{ROGUE_TBL}
\end{table}

\section{Feshbach Tuning STIRAP}
\label{FESHTUNE}

As we have already seen, reducing the density of the initial atomic condensate
acts to marginalize collisional mean-field shifts of the two-photon resonance,
leading to an improvement in the efficiency of STIRAP from an atomic to a
molecular condensate. With this idea in mind, it seems possible that
collisional mean-field shifts could be further marginalized by adjusting the
atom-atom scattering length via off-resonant magnetoassociation. We address
this problem by first rewriting the Hamiltonian~\eq{H_0} in a form that makes
the mechanics of the Feshbach-tuned efficiency more transparent.

Neglecting rogue photodissociation to noncondensate modes as per the previous
section, consider now the total number of particles, which is given by the
operator
\beq
N=\adag a+2(\bdag b+\gdag g).
\eeq
Due to the damping term ($\propto i\gamma_s$) in the Hamiltonian~\eq{H_0},
it cannot be said that $N$ is a conserved quantity. Nevertheless, it is
certainly true that $N$ commutes with the Hamiltonian, i.e., $[H,N]=0$,
regardless of the non-Hermitian nature of $H$. It is then reasonable to
presume that one may add multiples of $N$ to the Hamiltonian without
altering the essential physics. For example, it is implicitly accepted
that the two free Hamiltonians $H_0/\hbar=\half\Delta\adag a+\delta\bdag b$ and
$H_0/\hbar=-\Delta\gdag g+(\delta-\Delta)\bdag b$ will give the same
physics, and the two are of course related by the
addition (subtraction) of the term $\half\hbar\Delta N$ [see also derivation
of Eqs.~\eq{DARKY}]. Similarly, it is also reasonable to expect that subtracting
the term
$\half\hbar\lambda_{11}N^2$ will leave the dynamics unchanged, an operation that
results in the Hamiltonian
\bea
\frac{H}{\hbar}&=&\half\Delta \adag a + \delta\bdag b
   + \half\lambda_{33}' \gdag\gdag gg
     +\lambda_{13}'\adag\gdag ag,
\nonumber\\
   &&-\half\left[\kappa(\bdag aa + \adag\adag b)
     +\Omega(\gdag b + \bdag g)\right],
\label{H_PRI}
\eea
where the
parameters $\lambda_{33}'=\lambda_{33}-4\lambda_{11}$ and
$\lambda_{13}'=\lambda_{13}-2\lambda_{11}$ now characterize the
mean-field shifts
due to collisional interactions, and we continue to neglect the mean-field
shifts associated with electronically-excited molecules.

The reduced-parameter
Hamiltonian~(\ref{H_PRI}) gives the mean-field equations
\bml
\bea
i\dot{a} &=& \left(\half\Delta+\Lambda_{13}'|g|^2\right)a
   - \chi a^* b, \\
i\dot{b} &=& \left(\delta-\half i\gamma_s \right) b
   -\half\left(\chi aa + \Omega g\right), \\
i\dot{g} &=& \left(\Lambda_{13}' |a|^2 + \Lambda_{33}'|g|^2\right) g
   -\half\Omega b.
\eea
\label{MFE_PRI}
\eml
We have made an extensive numerical comparison between the solutions to
reduced-parameter mean-field model~\eq{MFE_PRI} and the solutions to the
conventional mean-field model~\eq{MFE_0}. As expected, the conventional
and reduced-parameter mean-field equations of motion lead to exactly the same
physics. Moreover, we have also confirmed that, barring a fluke in the
scattering length(s), collisions with the excited-state are safely ignored.

\begin{figure}
\centering
\includegraphics[width=8.0cm]{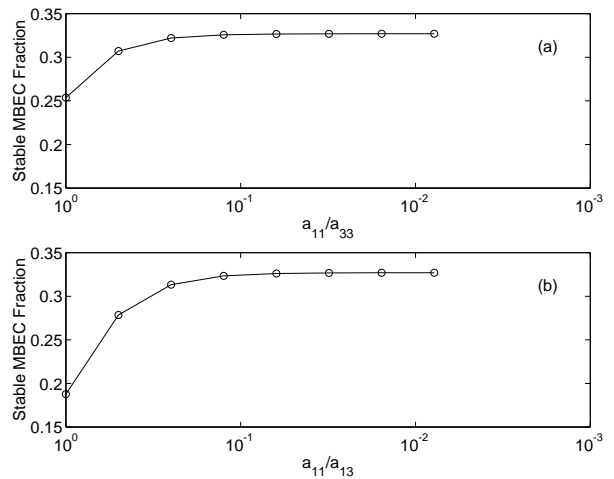}
\caption{Feshbach-tuning the efficiency of stimulated Raman adiabatic
passage in photoassociation of an interacting
condensate. Here the density is $\rho=4.3\times10^{12}\,{\rm
cm}^{-3}$, and the STIRAP parameters are the same as in
Fig.~\protect\ref{DTUNE}~(c) for asymmetric pulses. (a)~Compared to native
collisions ($a_{11}=a_{33}$), a Feshbach-tuned atomic scattering length offers
additional efficiency gain, which saturates at $a_{11}=a_{33}/8$. (b)~Similar
gain saturation also occurs at $a_{11}=3a_{13}/8<0$.}
\label{FTUNE}
\end{figure}

Besides reducing
the number of parameters, it seems that
$\Lambda'_{33}=0$, which corresponds to $a_{11}=a_{33}/8$, is a special
case that could further improve the molecular conversion efficiency. Indeed,
reducing the scattering length produces another marked gain in the molecular
conversion efficiency [Fig.~\ref{FTUNE}~(a)]. The gain saturates because
$\lambda_{11}$ is steadily decreasing, so that
$\lambda'_{33}\rightarrow\lambda_{33}$. Similar saturation is shown in
Fig.~\ref{FTUNE}~(b) for $\lambda_{13}'=0$, but any gain here is outweighed by
the fact that $a_{11}<0$ destabilizes the initial BEC against collapse, although
a system where $a_{13}>0$ may prove otherwise. Again, further improvements can
be achieved by optimizing the laser parameters~\cite{DRU02,HOR02}.

Before closing, we extend our model to explicitly describe tuning
the scattering length via magnetoassociation. Hence, in addition to the
primary-photoassociation molecular state, the BEC atoms are coupled to a second
molecular state, denoted by the operator
$\phi$, with a Feshbach resonance. For well-resolved resonances, light-driven
transitions involving the Feshbach state are neglected just as, for example,
direct free-bound photoassociation transitions are neglected in
Feshbach-stimulated photoproduction of molecular BEC~\cite{KOK01,MAC02}. 
The Hamiltonian is then appended to read
\beq
\frac{H}{\hbar} = \ldots+\tilde\omega_B\phidag\phi
    -\frac{\alpha}{\sqrt{N}}\left(\phidag aa + \adag\adag\phi\right),
\label{H_MA}
\eeq
where $\tilde\omega_B=[\omega_B-i(\Gamma_d+\Gamma_\rho\adag a)]$. Here
$\Gamma_d$ is the rate of magnetodissociation to noncondensate modes and
$\Gamma_\rho$ is the loss rate due to collision-induced vibrational relaxation;
also, the magnetoassociation coupling is
$\alpha=\sqrt{2\pi\rho|a_{11}|\Delta_\mu\Delta_B/m}$,
and the energy (detuning) of the magnetoassociation state is 
$\hbar\omega_B={\rm sgn}[a_{11}]\Delta_\mu(B-B_0)$, with $\Delta_\mu$ the
difference between the atom pair and molecule magnetic
moments and
$\Delta_B$ is the width of the resonance. Anticipating lowest-order
perturbation theory, we have neglected
molecule-molecule collisions.

When the system is off-resonance ($\omega_B\gg\alpha,\Gamma_d,\rho\Gamma_\rho$),
it is easily shown from the Heisenberg equation of motion for the molecular
operator that, to lowest order, off-resonant magnetoassociation can be included
simply by making the substitution
\beq
\Lambda_{11}\rightarrow\tilde\Lambda_{11}=\Lambda_{11}
  -\frac{2\alpha^2}{\omega_B}
    \left[1+i\left(\frac{\Gamma_d}{\omega_B}
      +\frac{\rho\Gamma_\rho}{\omega_B}\right)\right].
\eeq
According to Fig.~\ref{FTUNE}, we are looking at about an order of magnitude
decrease in the scattering length. For concreteness, we assume
$\Re[\tilde\Lambda_{11}]=\Lambda_{11}/10$, which translates into the magnetic
field position $B=B_0+0.9\Delta_B$. The question is then whether this is
sufficiently detuned from resonance to justify neglect of dissociative and
relaxational losses. The ultracold magnetodissociation rate is~\cite{YUR02}
$\Gamma_d=a_{11}\Delta_\mu\Delta_Bp/\hbar^2$, where $p$ is the relative
atom-pair momentum; hence, using the uncertainty relation
$\rho^{-1/3}p\sim\hbar/2$, we find $\Gamma_d/\omega_B\sim
a_{11}\rho^{1/3}\Delta_B/(B-B_0)\sim10^{-3}$. Estimating
$\Gamma_\rho\sim10^{-10}{\rm cm}^3/{\rm s}$~\cite{YUR02},
$\Delta_B\sim0.1\,\rm{G}$ and $\Delta_\mu=\mu_B$ (i.e., a Bohr magneton), then
the loss term due to vibrational quenching is roughly
$\rho\Gamma_\rho/\omega_B\sim10^{-4}$ for $\rho\sim10^{12}\,{\rm cm}^{-3}$. It
should therefore be possible, using one of the more than forty Feshbach
resonances found in $^{87}$Rb~\cite{VKE02}, to improve
the STIRAP efficiency by tuning the atom-atom scattering length.

\section{Discussion}
\label{DISC}

Admittedly, we have neglected an explicit trapping
potential for either the atoms or the molecules, a move which we now justify.
From Fig.~\ref{DTUNE}~(c), the timescale for STIRAP is
$T=5\times10^3/\chi_0\sim\,24\,{\rm ms}$. For $N=5\times10^5$ condensate atoms
in a spherically symmetric trap, let us say that the reduction in density from
$\rho_0=4.3\times10^{14}\,{\rm cm}^{-3}$ to $\tilde\rho\sim10^{12}\,{\rm
cm}^{-3}$ is achieved solely by reducing the frequency of the trap from
$\nu_t=100\,{\rm s}^{-1}$ to $\nu_t\sim1\,{\rm s}^{-1}$. The STIRAP timescale
is then fast enough to allow inclusion of the trap with an average over a
Thomas-Fermi density distribution~\cite{DAL99}, which should not drastically
alter the predicted optimal density.

Also, the Raman-formed molecules can be vibrationally
hot~\cite{DRU02}, and molecule-molecule and atom-molecule
collisions may thus foster relaxation to lower-lying vibrational levels. This
vibrational quenching is accounted for in exactly the same fashion as already
presented in Sec.~\ref{FTUNE} for the Feshbach molecular state. An exact value
for the $^{87}{\rm Rb}_2$ quenching rate (per unit density) is unknown (hence
the neglect in Ref.~\cite{DRU02}), although the upper bound
$\Gamma'_\rho<10^{-10}\,{\rm cm}^3/{\rm s}$ has been measured~\cite{WYN00}.
At this rate, vibrational quenching could well be an issue, even for the
optimal density $\tilde\rho\sim10^{12}\,{\rm cm}^{-3}$. However, a lower peak
bound-bound Rabi coupling ($\Omega_0/\chi_0=50$ compared to
$10^3$~\cite{DRU02}) means that, for the same bound-bound laser
intensities, it should be possible to target lower-lying levels. Because the
vibrational-relaxation rate decreases with decreasing binding
energy~\cite{BAL98}, targeting lower-lying vibrational levels should also lead
to slower relaxation rates. In the end, vibrational quenching should be
manageable.

The remaining concern is whether the photoassociation laser will overlap with the
expanded BEC cloud for user-friendly intensities. If not, then the cloudsize
could be reduced by adjusting the number of condensate atoms; additionally,
Feshbach-tuning the scattering length will also reduce the cloudsize. In
particular, if the density $\rho=4.6\times10^{12}\,{\rm
cm}^{-3}\approx\tilde\rho$ is achieved by lowering the number of atoms to
$N=10^4$, the scattering length to $\tilde{a}_{11}=a_{11}/8$, and the trap
frequency to $\nu_t=3\,{\rm s}^{-1}$, then we find that the cloudsize is only
barely increased from $R_0=8.9\,\mu{\rm m}$ to $R=11\,\mu{\rm m}$. By
optimizing the number of atoms, the trap frequency and atomic scattering length,
overlap between the BEC cloud and the photoassociation laser should be manageable
for user-friendly intensities.

Previously~\cite{KOS00}, we have argued that experimental Raman
photoassociation of a BEC~\cite{WYN00} is on the verge of coherent
conversion. The experiments in question are performed with continuous-wave
lasers and spontaneous decay from the primary photoassociation state is
managed with a large intermediate detuning. To enable coherent phenomena, the
two-photon Rabi frequency, $\chi\Omega/\delta$, must be of the order of the
two-photon spontaneous decay linewidth, $\Omega^2\gamma_s/2\delta^2$, a
condition that is not quite satisfied in the Wynar~\etal~\cite{WYN00}
experiments. On the other hand, atom-atom collisions are also an issue in
achieving continuous-wave coherence, and avoiding them by increasing the laser
intensity and decreasing the density could potentially bring rogue
photodissociation and laser-cloud overlap into play. Our opinion remains that
coherent continuous-wave conversion can only be achieved by a
difficult-to-impossible balance of laser intensity and condensate density. This
bottleneck is probably not unique to the $^{87}$Rb system.

The present work has examined {\em transient} stimulated Raman adiabatic passage
in photoassociation of a freely-interacting condensate in the presence of rogue
and radiative decay. We have shown that rogue
photodissociation  to noncondensate modes is generally negligible, and that
radiative decay, enhanced by  collisional mean-field shifts, can be managed by
adjusting the initial condensate density and the atom-atom scattering length.
In conclusion, a fully-optimized STIRAP scheme--i.e., where the detunings,
pulse-sequence parameters (heights, widths, spacing, and number of
sequences), particle number, trap parameters, and atomic scattering length are
optimized--offers probably the best chance for achieving coherent atom-molecule
conversion in photoassociation.

\begin{acknowledgments}
The authors gratefully acknowledge financial support from the Academy of
Finland (MM and KAS, project 206108), the Magnus Ehrnrooth Foundation
(AC), as well as NSF and NASA (JJ, PHY-0097974 and NAG8-1428).
\end{acknowledgments}

\end{document}